# Determination of the Spin-Hall-Effect-Induced and the Wedged-Structure-Induced Spin Torque Efficiencies in Heterostructures with Perpendicular Magnetic Anisotropy

Chi-Feng Pai[1], Maxwell Mann, Aik Jun Tan, and Geoffrey S. D. Beach[2]

*Department of Materials Science and Engineering, Massachusetts Institute of Technology, Cambridge,*

*Massachusetts 02139, USA*

We report that by measuring current-induced hysteresis loop shift versus in-plane bias magnetic field, the spin Hall effect (SHE) contribution of the current-induced effective field per current density, $\chi_{\text{SHE}}$, can be estimated for Pt and Ta-based magnetic heterostructures with perpendicular magnetic anisotropy (PMA). We apply this technique to a Pt-based sample with its ferromagnetic (FM) layer being wedged-deposited and discover an extra effective field contribution, $\chi_{\text{Wedged}}$, due to the asymmetric nature of the deposited FM layer. We confirm the correlation between $\chi_{\text{Wedged}}$ and the asymmetric depinning process in FM layer during magnetization switching by magneto-optical Kerr (MOKE) microscopy. These results indicate the possibility of engineering deterministic spin-orbit torque (SOT) switching by controlling the symmetry of domain expansion through the materials growth process.

---

[1] Current address: Department of Materials Science and Engineering, National Taiwan University, Taipei 10617, Taiwan
[2] gbeach@mit.edu

Current-induced spin-orbit torque (SOT) has been shown to be an efficient way of manipulating the magnetization in heavy-metal/ferromagnet (HM/FM) heterostructures. Unlike conventional spin transfer torque [1, 2], in which the source of spin angular momentum comes from a ferromagnetic polarizer layer, SOTs arise from either the bulk-like spin Hall effect (SHE) [3, 4] of the nonmagnetic HM layer or Rashba-type spin-orbit interaction [5, 6] at the interface. SOTs can be utilized to achieve efficient magnetization switching [7-10], ultrafast domain wall (DW) motion [6, 11, 12], and microwave generation through magnetic oscillations [13, 14] in spintronic device applications.

SOTs are typically studied in magnetic heterostructures with perpendicular magnetic anisotropy (PMA), and in general both a Slonczewski-like and a field-like torque can be present. The Slonczewski-like torque is most relevant to magnetization switching: it manifests as an effective field $H_{\text{eff}}$ with an out-of-plane (easy-axis) component that can reverse the magnetization or drive DWs if a component of the magnetization lies along the current-flow direction. The most common measurement schemes used to quantify the Slonczewski-like SOT efficiency $\chi \equiv H_{\text{eff}} / J_e$ (effective field per unit current density $J_e$) include ferromagnetic resonance techniques [15-17], low-frequency harmonic voltage measurements using small AC currents [18-20], and analysis of current-induced DW motion in thin magnetic strips [11, 12, 21, 22]. Current-induced SOT switching of PMA films under an in-plane bias field is another convenient means for determining the sign of χ; however, a quantitative estimate of its magnitude is usually difficult to obtain in such measurements due to the complicated magnetization reversal process [23, 24].

In this work we examine the role of domain nucleation and DW propagation in SOT-assisted magnetization switching in HM/FM bilayer systems with PMA. We show that the current-induced shift of the out-of-plane hysteresis loop as a function of in-plane bias field can be well-explained by a simple current-assisted DW propagation model. This simple measurement scheme allows $\chi$ to be quantified and simultaneously provides an estimate of the chiral Dzyaloshinskii-Moriya effective field $|H_{\rm DMI}|$ that stabilizes Neel-type DWs in these structurally inversion asymmetric structures. Finally, we show that in wedged films with a small thickness gradient there exists a large apparent contribution $\chi_{\rm Wedged}$ to the Slonczewski-like SOT efficiency that derives from structural asymmetries in the domain nucleation/propagation process, which is further examined by magneto-optical Kerr (MOKE) microscopy. Importantly, we find that this effect can provide a means for deterministic SOT switching of a PMA film in the absence of an in-plane bias field. This result may offer an alternate explanation to similar recently-reported observations interpreted in terms of an out-of-plane effective field generated by in-plane symmetry breaking [25].

In magnetic heterostructures with PMA, the SHE-induced Slonczewski-like SOT can drive Neel DWs similarly to an out-of-plane applied field, in a direction that depends on the DW chirality. Homochiral Neel DWs can be stabilized by the Dzyaloshinskii-Moriya interaction (DMI) in ultrathin films lacking inversion symmetry [11, 12, 21, 26], and it has been shown that current-induced magnetization switching and DW motion in HM/FM bilayers can be explained by a SHE+DMI scenario [11, 23]. As schematically shown in Fig. 1(a), the charge current $J_e$ flowing along x-axis in the NM layer of a magnetic heterostructure will generate a transverse spin current $J_s$ along z-axis via the SHE and inject spins into FM layer with their spin-polarization direction $\hat{\sigma}$ parallel to y-axis. This spin current, when acting upon a Neel-type DW with the typical Walker profile, will give rise to an effective field $H_{\rm eff}^z = \chi J_e$ [27] through spin transfer torque mechanism, where

$\chi = \chi_{\rm SHE} \cos\Phi = (\pi/2)(\hbar \xi_{DL} / 2e\mu_0 M_s t_{FM}) \cos\Phi$. Here $\xi_{DL}$, $M_s$, $t_{FM}$, and $\Phi$ represent

the effective spin-Hall-induced (damping-like) torque efficiency [28, 29], the saturation magnetization of the FM, the thickness of FM, and the angle between DW moment and x-axis, respectively. In the case of homochiral Neel DWs, this $H_{\text{eff}}^z$ can lead to DW motion but *not* domain expansion in the absence of external magnetic field due to the opposite signs of $H_{\text{eff}}^z$ for up-down ($\cos\Phi = 1$) and down-up ($\cos\Phi = -1$) DWs. However, upon the application of an in-plane bias field $H_x$ that is strong enough to overcome the effective DMI field $H_{\text{DMI}}$, the DW moment in the Neel-type walls will re-align parallel to $H_x$ as shown in Fig. 1(b). In this case $H_{\text{eff}}^z$ will be pointing along the same direction for both up-down and down-up walls and therefore facilitates domain expansion or contraction, depending on the polarities of $J_e$ and $H_x$. Field-driven magnetization switching should therefore depend on both $J_e$ and $H_x$. It is then straightforward to conceive that not only for *current*-driven DW motion and/or magnetization switching, but also for an out-of-plane *field*-driven switching process, the applied current $J_e$ and in-plane bias field $H_x$ should play significant roles.

To study the interplay between $J_e$, $H_x$, and resulting $H_{\text{eff}}^z$ during field-driven switching, we prepared four sets of PMA Hall-bar samples: (A) ||Ta(2)/Pt(4)/Co(1)/MgO(2)/Ta(1), (B) ||Ta(2)/Pt(4)/CoFeB(1)/MgO(2)/Ta(1), (C) ||Ta(6)/CoFeB(1)/MgO(2)/Ta(1), and (D) ||Ta(2)/Pt(4)/Co($t_{\text{Co}}$)/MgO(2)/Ta(1) with $0.6\,\text{nm} \leq t_{\text{Co}} \leq 1.6\,\text{nm}$ being wedged-deposited. || stands for thermally-oxidized Si substrate and the numbers in parenthesis represent nominal thickness of sputtered films in nanometers. All films were sputter-deposited in an AJA ATC-series sputtering chamber with base pressure $\leq 10^{-7}$ Torr and a working Ar pressure of 4 mTorr. The substrate-to-target separation was $\approx 15\,\text{cm}$ with an oblique angle, and the uniform thickness of films was achieved by substrate rotation during deposition. The wedged-deposition of the Co layer for series (D) was achieved by sputtering with the rotation function off. Hall bars with lateral dimensions of 5 μm by 12 μm were patterned using standard photolithography, and Ti(5)/Pt(50) pads were deposited by sputtering for electrical contact.

As schematically shown in Fig. 2(a), we measured the anomalous Hall (AH) voltage $V_H$ vs out-of-plane field $H_z$ to characterize magnetization switching in the Hall-bar devices, as a function of applied DC current $I_{DC}$ and in-plane bias field $H_{\text{in-plane}}$ (either along the x-axis or y-axis). Representative normalized AH loops for Pt(4)/Co(1)/MgO(2) sample (A) with $H_x = 2500\,\text{Oe}$ and $I_{DC} = \pm 6\,\text{mA}$ are shown in Fig. 2(b). Slight vertical offsets are introduced for both AH loops for clarity. The opposite loop shifts along the $H_z$-axis of the hysteresis loops corresponding to opposite polarities of $I_{DC}$ indicate the existence of a current-induced $H_{\text{eff}}^z$ due to a Slonczewski-like torque. By plotting the switching fields $H_{SW}$ for both down-to-up and up-to-down transitions (defined as the zero-crossing fields of the normalized $V_H$) as functions of $I_{DC}$, we obtained a switching phase diagram as shown in Fig. 2(c). Two effects should be considered to explain the variation of switching boundaries: linear tilting contribution from current-induced $H_{\text{eff}}^z$ and the reduction of coercivity from Joule heating. However, the Joule heating contribution can be eliminated by considering only on the horizontal shift of the hysteresis loop center $H_0$, defined as the mean of the two switching fields $\left(H_{SW}^{\text{DN-to-UP}} + H_{SW}^{\text{UP-to-DN}}\right)/2$. The linear variation of $H_0$ with respect to $I_{DC}$ then provides a good estimate of $H_{\text{eff}}^z / I_{DC}$. For comparison, we also plot the results measured from sample (A) in the absence of $H_x$ in Fig. 2(d). As expected, no contribution other than Joule heating was observed since the Neel-type DWs were not re-aligned in order to affect the domain expansion/contraction processes.

To verify this measured $H_{\text{eff}}^z / I_{DC}$ is indeed coming from the SHE, we performed the same measurements on sample (B) and (C), namely PMA Pt(4)/CoFeB(1)/MgO(2) and Ta(6)/CoFeB(1)/MgO(2). The only difference between these two samples is the choice of NM underlayer that is the source of the SHE. In Fig. 3(a) we plot the representative AH loops for Pt/CoFeB/MgO and a similar shift to $I_{DC}$ as in Pt/Co/MgO was found. The $I_{DC}$ dependence of the measured $H_{\text{eff}}^z$ for Pt/CoFeB/MgO is summarized in Fig. 3(b). It can be seen that by reversing the polarity of $H_x$, the slope of $H_{\text{eff}}^z / I_{DC}$ also reverses. This is consistent with the prediction from a SHE+DMI scenario.

Finally, results from the Ta/CoFeB/MgO sample are shown in Fig. 3(c) and (d). An opposite trend of $H_{\text{eff}}^z / I_{DC}$ was found compare to the Pt case. Since Pt and Ta are known to have opposite spin Hall angles [8], this opposite trend of $H_{\text{eff}}^z / I_{DC}$ between the two cases is again consistent with the SHE picture.

In Fig. 4(a), (b), and (c), we summarize the measured $\chi = H_{\text{eff}}^z / J_e$ as a function of applied in-plane field, either along the x-axis or y-axis, for sample (A) Pt/Co/MgO, sample (B) Pt/CoFeB/MgO, and sample (C) Ta/CoFeB/MgO, respectively. The current density in the NM layer was calculated from $I_{DC}$, dimensions of the Hall-bar device, and resistivities of the NM/FM layers. For Pt/Co/MgO (Fig. 4(a)), $\chi$ increases quasi-linearly with $H_x$ and saturates at $H_x \approx 5000\,\text{Oe}$, while no significant trend or variation of $\chi$ was observed with the application of $H_y$. This observation is consistent with a domain-expansion picture: the DW orientations in the heterostructure change from an average of $\langle\cos\Phi\rangle \approx 0$ to $\langle\cos\Phi\rangle \approx 1$ when $H_x$ approaches $H_{\text{DMI}}$, whereas $H_y$ simply re-orients the DWs into a Bloch-type configuration ($\Phi = \pm\pi/2$). Based on this model, we estimated $\chi_{\text{SHE}} \approx 75\,\text{Oe}/10^{11}\text{A/m}^2$ and $|H_{\text{DMI}}| \approx 5000\,\text{Oe}$ for Pt/Co/MgO from saturation value of $\chi$ and the saturation field, respectively, in Fig. 4(a). For the Pt/CoFeB/MgO sample, we found $\chi_{\text{SHE}} \approx 72\,\text{Oe}/10^{11}\text{A/m}^2$ and $|H_{\text{DMI}}| \approx 2500\,\text{Oe}$ (Fig. 4(b)). The close correspondence of $\chi_{\text{SHE}}$ between Pt/CoFeB/MgO and Pt/Co/MgO case is expected since the spin Hall metal is the same. However, the significant difference in the DMI effective field for CoFeB and Co-based structures suggests that the exchange interaction is particularly sensitive to the FM layer composition. Finally, for the Ta/CoFeB/MgO sample, in Fig. 4(c), we estimated $\chi_{\text{SHE}} \approx -50\,\text{Oe}/10^{11}\text{A/m}^2$ and $|H_{\text{DMI}}| \approx 250\,\text{Oe}$. Note that the estimated $|H_{\text{DMI}}| \approx 2500\,\text{Oe}$ for Pt/CoFeB/MgO and $|H_{\text{DMI}}| \approx 250\,\text{Oe}$ for Ta/CoFeB/MgO are comparable to the previously reported values of $|H_{\text{DMI}}| \approx 2800\,\text{Oe}$ for Pt/CoFe/MgO and $|H_{\text{DMI}}| \approx 80\,\text{Oe}$ for Ta/CoFe/MgO structures [22].

In Fig. 4(d) we plot $|H_{\text{DMI}}|$ for these samples against their measured perpendicular anisotropy fields $H_{an}$. We find that for the Ta/CoFeB/MgO sample, the DMI is just beyond the threshold required to stabilize Neel DWs, given by $|H_{\text{DMI}}|/H_{an} = 2/\pi$ [27]. This indicates the possibility of stabilizing two-dimensional spin textures such as skyrmions in Ta-based magnetic heterostructure as well as its variations [30, 31].

The effective damping-like torque efficiencies (effective spin Hall angles) corresponding to the measured $\chi_{SHE}$ are $\xi_{DL} \approx 0.15$ and $\xi_{DL} \approx -0.12$ for Pt and Ta samples, respectively. These numbers are in good agreement with other recently reported values that were obtained through harmonic voltage measurements [20, 28] and spin-torque switching measurements [8, 29]. Moreover, the magnitude of the DMI exchange constant $|D|$ can be calculated from the measured $|H_{DMI}|$ by using $|D| = \mu_0 M_s \Delta |H_{DMI}|$ [27], where $\Delta$ is the DW width and relates to exchange stiffness constant $A$ and effective PMA energy density $K_{u,eff}$ through $\Delta = \sqrt{A/K_{u,eff}}$. Using $M_s$ and $K_{u,eff}$ obtained by vibrating sample magnetometry, and by assuming $A \approx 1.5 \times 10^{-11}$ J/m [32], we estimated $|D| \approx 3.0$ mJ/m$^2$ for Pt/Co/MgO, $|D| \approx 1.8$ mJ/m$^2$ for Pt/CoFeB/MgO, and $|D| \approx 0.6$ mJ/m$^2$ for Ta/CoFeB/MgO samples, respectively. Again, these numbers are reasonable and close to the range of previously reported values in similar magnetic heterostructure systems [22, 26, 33].

Recently it has also been shown that by engineering the gradient of $H_{an}$ in deposited films [25] or by artificially tilting the anisotropy axis away from the film normal [34], deterministic SOT switching in the absence of external magnetic field can be realized for Ta/CoFeB/Oxide heterostructures. Similar effect is also observed in structures with NM layer being TaN, Hf, or W and is attributed to the details during asymmetric materials growth [35]. However, this additional contribution to the SOT has not yet been reported for Pt-based magnetic heterostructures. Here we show that, by performing the same measurements as previous sections on sample (D) Pt/Co(wedge)/MgO, an additional contribution to $\chi$ is observed, and found to originate from the nature of the wedged structure, $\chi_{Wedged}$. We further show that this contribution can be quantified and distinguished from $\chi_{\text{SHE}}$ using the measurement scheme described above.

As shown in Fig. 5(a), unlike samples (A)-(C), sample (D) before patterning has a wedged-deposited Co layer $0.6\,\text{nm} \leq t_{\text{Co}} \leq 1.6\,\text{nm}$. After patterning, the wedged profile is along the y-axis of the device (Fig. 1(a)), with only a slight variation of Co thickness ($\leq 10^{-3}$ nm) for each Hall-bar device. Note that unlike in Ref. [25], here the wedged-deposited layer is the FM rather than the capping layer. As shown in Fig. 5(b), the measured $\chi$ has non-zero offsets at $H_x = 0\,\text{Oe}$ for both $\langle t_{\text{Co}} \rangle = 0.65\,\text{nm}$ and $\langle t_{\text{Co}} \rangle = 1.09\,\text{nm}$ samples. Here $\langle \ \rangle$ represents the averaged nominal thickness of the measured device. This extra SOT contribution in the absence of in-plane applied field is significantly different from the uniformly-deposited case. We denote this offset as the contribution from the wedged structure, $\chi_{\text{Wedged}}$. More importantly, the sign of $\chi_{\text{Wedged}}$ can change depending on $\langle t_{\text{Co}} \rangle$, which is similar to the dependence of $H_{\text{eff}}^z$ on $\langle t_{\text{TaO}_x} \rangle$ reported in Ref. [25]. By increasing the applied field to $H_x \geq |H_{\text{DMI}}|$, the measured $\chi$ again saturates, with $\chi \approx \chi_{\text{SHE}} + \chi_{\text{Wedged}}$. The resulting trend allows us to unambiguously determine $\chi_{\text{SHE}}$, $\chi_{\text{Wedged}}$, as well as $|H_{\text{DMI}}|$ for these wedged-deposited samples through this simple protocol.

We summarize the measured $\chi_{\text{SHE}}$ and $\chi_{\text{Wedged}}$ of sample (D) as a function of $\langle t_{\text{Co}} \rangle$ in Fig. 5(c). $\chi_{\text{SHE}}$ reaches its maximum at $\langle t_{\text{Co}} \rangle = 1.09\,\text{nm}$ and is close to that of the uniformly-deposited case (sample(A)), while the magnitude of $\chi_{\text{Wedged}}$ reaches its extreme values at $\langle t_{\text{Co}} \rangle = 0.65\,\text{nm}$ with $\chi_{\text{Wedged}} \approx 10\,\text{Oe}/10^{11}\,\text{A/m}^2$ and $\langle t_{\text{Co}} \rangle = 1.22\,\text{nm}$ with $\chi_{\text{Wedged}} \approx -14\,\text{Oe}/10^{11}\,\text{A/m}^2$. This indicates that there is no direct correlation between the two

SOT contributions and it is therefore possible that they can be independently tuned through interfacial and structural engineering. We also note that the maximum magnitude of $|\chi_{\text{Wedged}}| \approx 14\,\text{Oe}/10^{11}\,\text{A/m}^2$ for the wedged-deposited Pt/Co/MgO samples presented here is comparable to the reported values for Ta/CoFeB/MgO and Ta/CoFeB/TaO$_x$ systems [25, 36]. However, current-induced switching in the absence of external field cannot be demonstrated with $|I_{DC}| \leq 8\,\text{mA}$ (corresponding to $H_{\text{eff}}^z \leq 40\,\text{Oe}$) due to the large coercivity ($H_c \geq 200\,\text{Oe}$) of the present films. Further materials engineering to reduce $H_c$ in Pt/Co/MgO heterostructures while maintaining high $|\chi_{\text{Wedged}}|$ should allow for realizing deterministic current-induced switching without any external field in this Pt-based heterostructure.

To gain insights on the microscopic origin of the structural-induced $\chi_{\text{Wedged}}$, we study the magnetization switching process in both uniformly-deposited sample (A) and wedged-deposited sample (D) through wide-field MOKE microscopy. As shown in Fig. 5(d), we found that by applying an out-of-plane field $H_z$, the preferred nucleation sites are very different for devices from sample (A) (uniform Co) and from sample (D) (wedged Co). For devices with a uniform Co layer, the domains randomly nucleate at all edges and then further expand to accomplish magnetization reversal. For devices with a wedged Co layer (in this case a sample with $\langle t_{\text{Co}} \rangle = 1\,\text{nm}$), however, the nucleation processes for both up-to-down and down-to-up transitions always begin at the lower edge (thicker Co side) then the DW propagates across the device to the other edge (thinner Co side). It is surprising that this asymmetric nucleation/depinning process can be so significant even for a device with FM thickness variation $\leq 10^{-3}\,\text{nm}$. Since the variation in Co thickness across the wedge is vanishingly small, we speculate that this effect is due to an in-plane anisotropy component from the angled deposition that tends to align domain walls preferentially along a preferred axis. This preferred nucleation on one edge of the device is drastically different from other reports [25, 35, 36], in which no observable asymmetric field-driven nucleation process is found. We believe that, although the existence of $dH_{an}/dy$ gradient-induced field-like SOT [25] cannot be completely ruled out, a major contribution of the measured $\chi_{\text{Wedged}}$ originates from the asymmetric nature of nucleation/depinning process in these wedged-deposited Pt/Co/MgO devices. If this is the case, then further studies on the interplay among structural factors, especially the DMI [37] and the current-induced (Oersted) field [38-40] at the edges in these magnetic heterostructures, will be beneficial for engineering SOT switching without external bias field.

In summary, we demonstrate that by characterizing the shift of out-of-plane hysteresis loops under different DC currents and in-plane bias fields, the SOT efficiency $\chi = H_{\text{eff}}^z / J_e \approx \chi_{\text{SHE}}$ for Pt/Co/MgO, Pt/CoFeB/MgO, and Ta/CoFeB/MgO heterostructures can be obtained. The effective DMI field $H_{\text{DMI}}$ in above heterostructures can also be estimated simultaneously by this straightforward protocol. We can also estimate the extra contribution of $\chi$ due to bilateral-symmetry-breaking of the wedged-deposited FM layer in Pt/Co(wedge)/MgO, $\chi_{\text{Wedged}}$, with the same method. This together with the Kerr microscopic observation of the asymmetric domain nucleation in Pt-based wedged heterostructures provide insightful information on the roles of preferred nucleation sites and nucleation mode in engineering towards SOT switching in the absence of external fields.

**Acknowledgements**

The authors would like to acknowledge support from C-SPIN, one of the six SRC STARnet

Centers, sponsored by MARCO and DARPA. C.F. Pai would like to thank Dr. Kohei Ueda, Minh-Hai Nguyen, and Yongxi Ou for their comments and suggestions on the manuscript.

**References**

[1] J. C. Slonczewski, J. Magn. Magn. Mater. **159**, L1 (1996).

[2] L. Berger, Physical Review B **54**, 9353 (1996).

[3] M. I. Dyakonov and V. I. Perel, Phys. Lett. A **35**, 459 (1971).

[4] J. E. Hirsch, Phys. Rev. Lett. **83**, 1834 (1999).

[5] I. M. Miron, G. Gaudin, S. Auffret, B. Rodmacq, A. Schuhl, S. Pizzini, J. Vogel, and P. Gambardella, Nat. Mater. **9**, 230 (2010).

[6] I. M. Miron *et al.*, Nat. Mater. **10**, 419 (2011).

[7] I. M. Miron *et al.*, Nature **476**, 189 (2011).

[8] L. Q. Liu, C.-F. Pai, Y. Li, H. W. Tseng, D. C. Ralph, and R. A. Buhrman, Science **336**, 555 (2012).

[9] C.-F. Pai, L. Q. Liu, Y. Li, H. W. Tseng, D. C. Ralph, and R. A. Buhrman, Appl. Phys. Lett. **101**, 122404 (2012).

[10] L. Q. Liu, O. J. Lee, T. J. Gudmundsen, D. C. Ralph, and R. A. Buhrman, Phys. Rev. Lett. **109**, 096602 (2012).

[11] S. Emori, U. Bauer, S. M. Ahn, E. Martinez, and G. S. D. Beach, Nat. Mater. **12**, 611 (2013).

[12] K. S. Ryu, L. Thomas, S. H. Yang, and S. Parkin, Nat. Nanotech. **8**, 527 (2013).

[13] L. Q. Liu, C.-F. Pai, D. C. Ralph, and R. A. Buhrman, Phys. Rev. Lett. **109**, 186602 (2012).

[14] V. E. Demidov, S. Urazhdin, H. Ulrichs, V. Tiberkevich, A. Slavin, D. Baither, G. Schmitz, and S. O. Demokritov, Nat. Mater. **11**, 1028 (2012).

[15] K. Ando, S. Takahashi, K. Harii, K. Sasage, J. Ieda, S. Maekawa, and E. Saitoh, Phys. Rev. Lett. **101**, 036601 (2008).

[16] O. Mosendz, J. E. Pearson, F. Y. Fradin, G. E. W. Bauer, S. D. Bader, and A. Hoffmann, Phys. Rev. Lett. **104**, 046601 (2010).

[17] L. Q. Liu, T. Moriyama, D. C. Ralph, and R. A. Buhrman, Phys. Rev. Lett. **106**, 036601 (2011).

[18] U. H. Pi, K. W. Kim, J. Y. Bae, S. C. Lee, Y. J. Cho, K. S. Kim, and S. Seo, Appl. Phys. Lett. **97**, 162507 (2010).

[19] J. Kim, J. Sinha, M. Hayashi, M. Yamanouchi, S. Fukami, T. Suzuki, S. Mitani, and H. Ohno, Nat. Mater. **12**, 240 (2013).

[20] K. Garello *et al.*, Nat. Nanotech. **8**, 587 (2013).

[21] P. P. J. Haazen, E. Mure, J. H. Franken, R. Lavrijsen, H. J. M. Swagten, and B. Koopmans, Nat. Mater. **12**, 299 (2013).

[22] S. Emori, E. Martinez, K. J. Lee, H. W. Lee, U. Bauer, S. M. Ahn, P. Agrawal, D. C. Bono, and G. S. D. Beach, Phys. Rev. B **90**, 184427 (2014).

[23] O. J. Lee, L. Q. Liu, C. F. Pai, Y. Li, H. W. Tseng, P. G. Gowtham, J. P. Park, D. C. Ralph, and R. A. Buhrman, Phys. Rev. B **89**, 024418 (2014).

[24] G. Q. Yu, P. Upadhyaya, K. L. Wong, W. J. Jiang, J. G. Alzate, J. S. Tang, P. K. Amiri, and K. L. Wang, Phys. Rev. B **89**, 104421 (2014).

[25] G. Q. Yu *et al.*, Nat. Nanotech. **9**, 548 (2014).

[26] J. Torrejon, J. Kim, J. Sinha, S. Mitani, M. Hayashi, M. Yamanouchi, and H. Ohno, Nat. Commun. **5**, 4655 (2014).

[27] A. Thiaville, S. Rohart, E. Jue, V. Cros, and A. Fert, Europhys. Lett. **100**, 57002 (2012).

[28] C. F. Pai, Y. X. Ou, L. H. Vilela-Leao, D. C. Ralph, and R. A. Buhrman, Phys. Rev. B **92**, 064426 (2015).

[29] M. H. Nguyen, C. F. Pai, K. X. Nguyen, D. A. Muller, D. C. Ralph, and R. A. Buhrman, Appl. Phys. Lett. **106**, 222402 (2015).

[30] W. J. Jiang *et al.*, Science **349**, 283 (2015).

[31] S. Woo *et al.*, arxiv:1502.07376 (2015).

[32] P. J. Metaxas, J. P. Jamet, A. Mougin, M. Cormier, J. Ferre, V. Baltz, B. Rodmacq, B. Dieny, and R. L. Stamps, Phys. Rev. Lett. **99**, 217208 (2007).

[33] A. Hrabec, N. A. Porter, A. Wells, M. J. Benitez, G. Burnell, S. McVitie, D. McGrouther, T. A. Moore, and C. H. Marrows, Physical Review B **90**, 020402 (2014).

[34] L. You, O. Lee, D. Bhowmik, D. Labanowski, J. Hong, J. Bokor, and S. Salahuddin, PNAS **112**, 10310 (2015).

[35] J. Torrejon, F. Garcia-Sanchez, T. Taniguchi, J. Sinha, S. Mitani, J. V. Kim, and M. Hayashi, Phys. Rev. B **91**, 214434 (2015).

[36] G. Q. Yu, L. T. Chang, M. Akyol, P. Upadhyaya, C. L. He, X. Li, K. L. Wong, P. K. Amiri, and K. L. Wang, Appl. Phys. Lett. **105** (2014).

[37] S. Pizzini *et al.*, Phys. Rev. Lett. **113**, 047203 (2014).

[38] X. Fan, H. Celik, J. Wu, C. Y. Ni, K. J. Lee, V. O. Lorenz, and J. Q. Xiao, Nat. Commun. **5**, 3042 (2014).

[39] D. Bhowmik, M. E. Nowakowski, L. You, O. Lee, D. Keating, M. Wong, J. Bokor, and S. Salahuddin, Scientific Reports **5**, 11823 (2015).

[40] C. J. Durrant, Q. Hao, G. Xiao, and R. J. Hicken, arXiv:1508.00833 (2015).

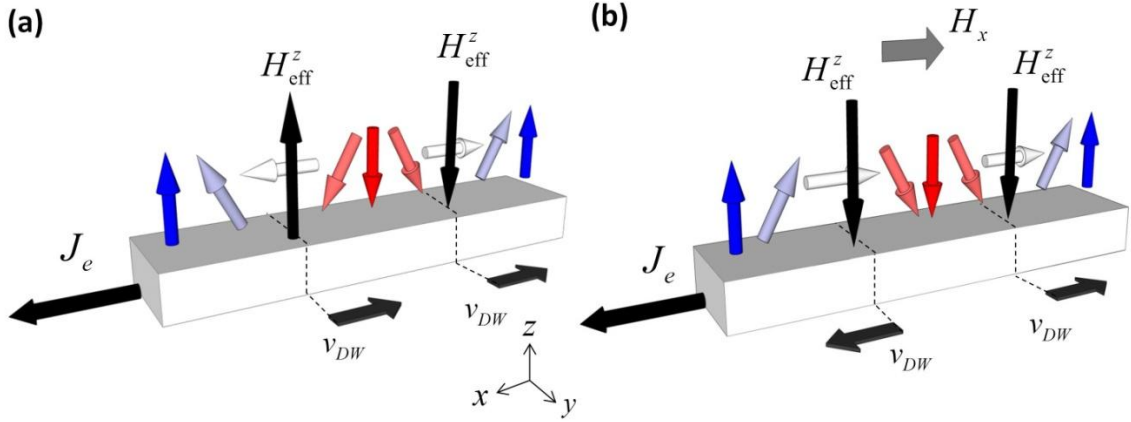

FIG. 1 (a) Schematic illustration of current-induced domain wall motion in a magnetic heterostructure with PMA in the absence of external magnetic field. $H_{\text{eff}}^z$ represents the SHE-induced effective field acting upon the Neel-type chiral domain wall. $v_{\text{DW}}$ represents the domain wall motion direction. (b) Schematic illustration of current-induced domain wall motion (domain expansion) with an in-plane external magnetic field $H_x$ to re-align domain wall moments.

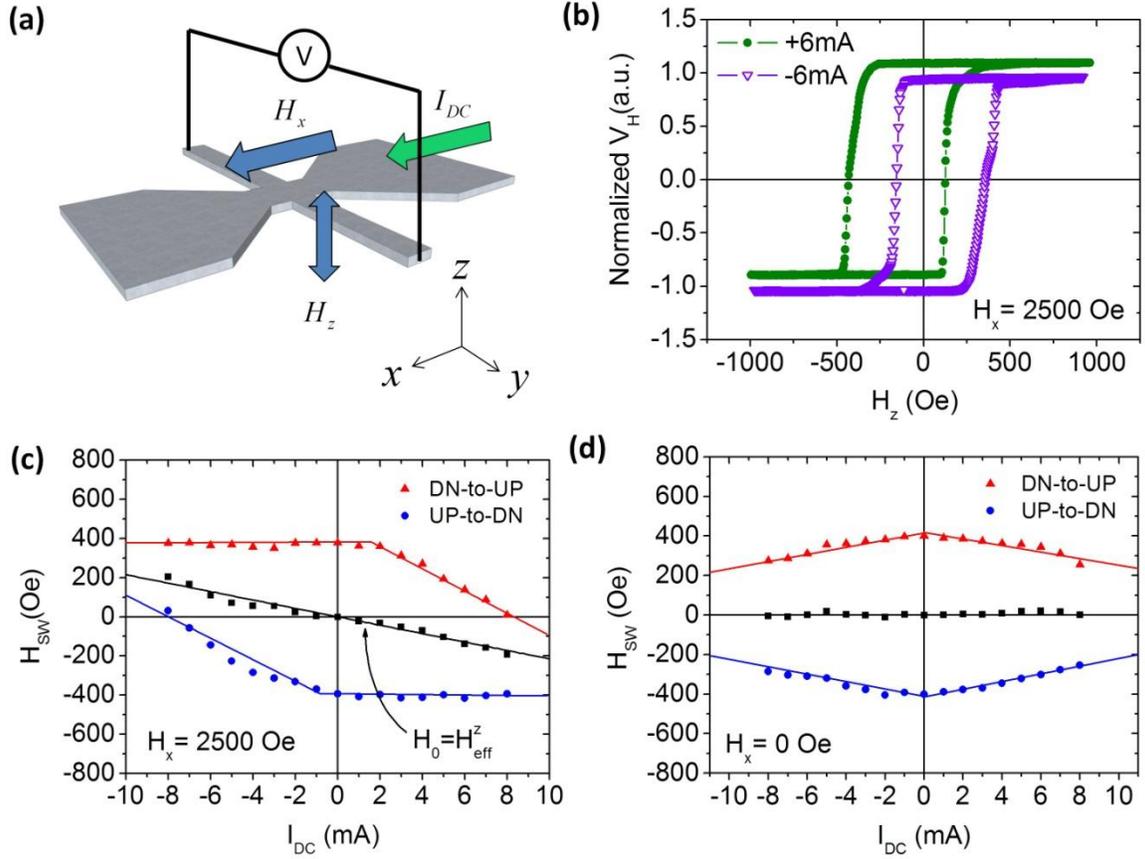

FIG. 2 (a) Schematic illustration of anomalous Hall (AH) voltage measurements. (b) AH loops for a Pt(4)/Co(1)/MgO(2) sample with DC currents $I_{DC} = \pm 6\,\text{mA}$ and an in-plane bias field $H_x = 2500\,\text{Oe}$. (c) Switching (depinning) fields $H_{SW}$ for down-to-up (red triangles) and up-to-down (blue circles) magnetization reversals as functions of $I_{DC}$, with $H_x = 2500\,\text{Oe}$. $H_0 = H_{\text{eff}}^z$ (black squares) represent the center of the AH loops. (d) $H_{SW}$ and $H_0$ as functions of $I_{DC}$ in the absence of in-plane bias field ($H_x = 0\,\text{Oe}$). Solid lines are guides for the eye.

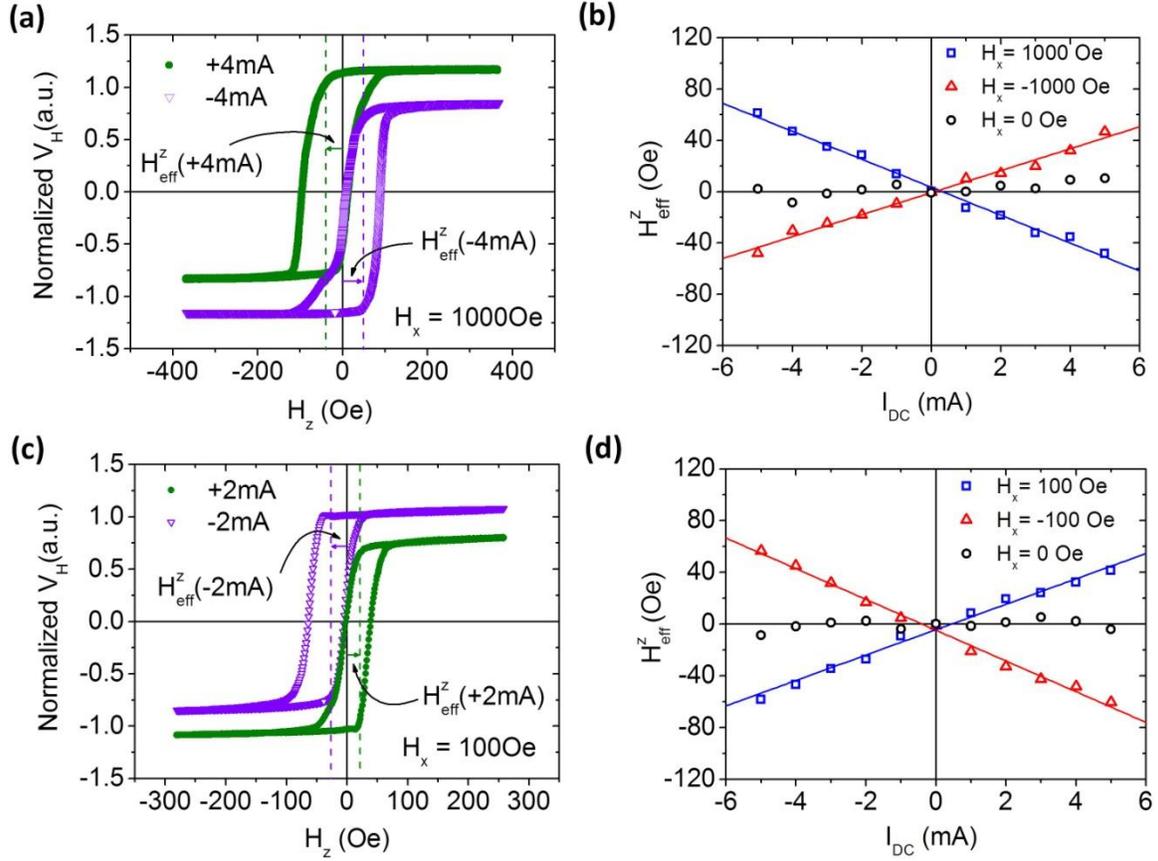

FIG. 3 (a) AH loops for a Pt(4)/CoFeB(1)/MgO(2) sample with DC currents $I_{DC}=\pm 4\,\text{mA}$ and an in-plane bias field $H_x=1000\,\text{Oe}$. $H_{\text{eff}}^z$ represents the shift of the AH loops due to the SHE. (b) $H_{\text{eff}}^z$ for Pt(4)/CoFeB(1)/MgO(2) as a function of $I_{DC}$ under different bias fields. (c) AH loops for a Ta(6)/CoFeB(1)/MgO(2) sample with DC currents $I_{DC}=\pm 2\,\text{mA}$ and an in-plane bias field $H_x=100\,\text{Oe}$. (d) $H_{\text{eff}}^z$ for Ta(6)/CoFeB(1)/MgO(2) as a function of $I_{DC}$ under different bias fields. Solid lines are linear fits to the data.

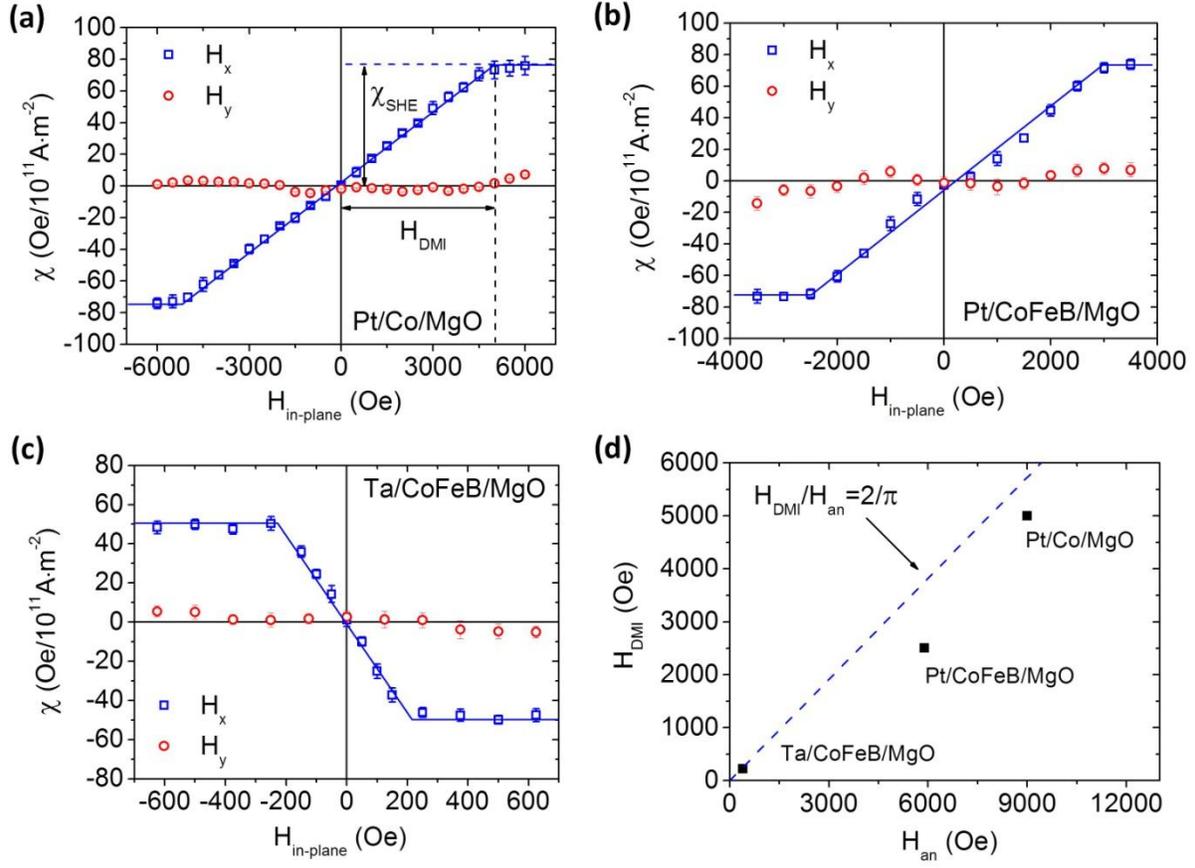

FIG. 4 The measured effective $\chi$ as a function of applied in-plane field for (a) Pt(4)/Co(1)/MgO(2), (b) Pt(4)/CoFeB(1)/MgO(2), and (c) Ta(6)/CoFeB(1)/MgO(2) samples. Blue squares and red circles represent data obtained with the external in-plane magnetic field applied along x-axis ($H_x$) and y-axis ($H_y$), respectively. Solid lines serve as guides for the eye. (d) The estimated DMI field $H_{DMI}$ as a function of anisotropy field $H_{an}$ for the presented samples. The dashed line represents the criterion above which skyrmions and other spin textures can be realized.

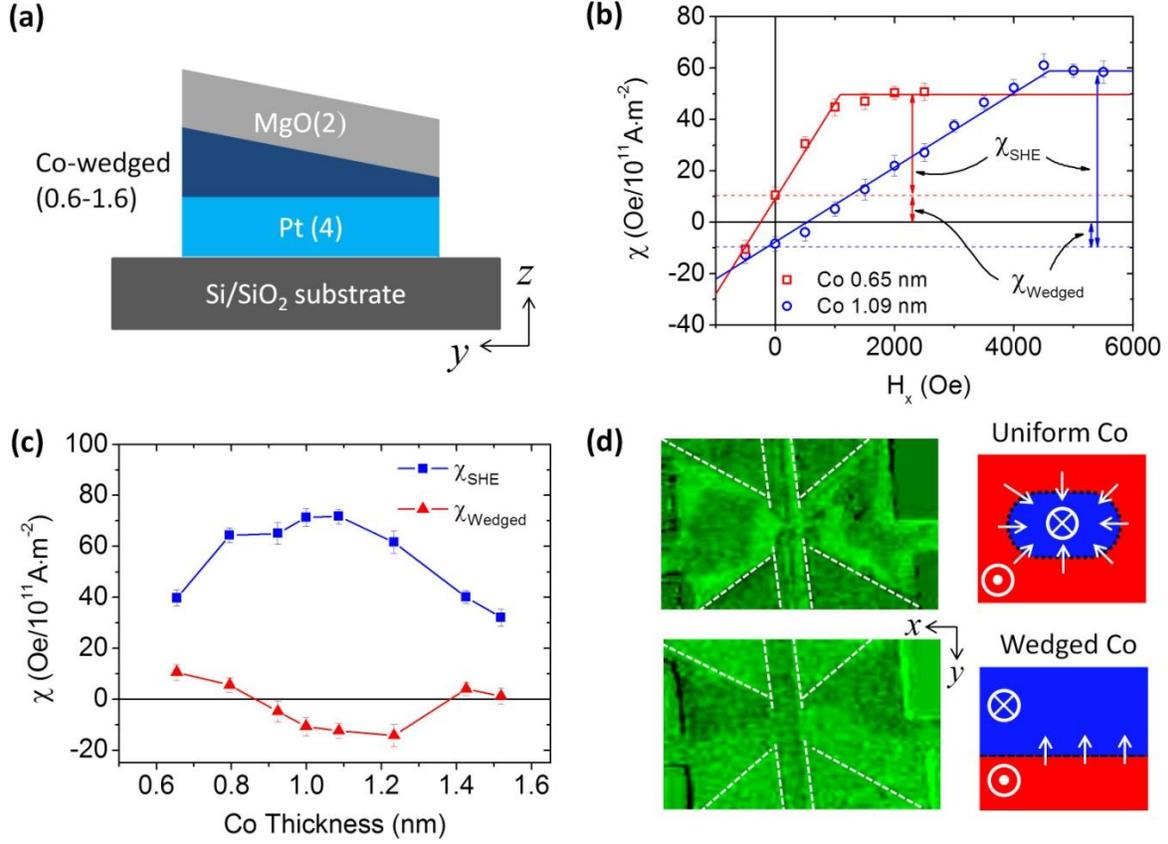

FIG. 5 (a) Schematic illustration of the wedged-deposited film. After patterning, the current was applied along into-the-plane direction during measurements in this cross-sectional view. (b) The measured $\chi$ as a function of $H_x$ for wedged-deposited Pt(4)/Co(0.65)/MgO(2) and Pt(4)/Co(1.09)/MgO(2) heterostructures. The contributions from the SHE and the wedged-structure are indicated as $\chi_{SHE}$ and $\chi_{Wedged}$, respectively. (c) Co thickness $\langle t_{Co} \rangle$ dependence of $\chi_{SHE}$ and $\chi_{Wedged}$. (d) Representative MOKE images for Pt(4)/Co(1)/MgO(2) devices during magnetization switching. The upper (lower) figures represent the switching mode for a uniformly (wedged)-deposited sample.